\documentclass[aps,prd,preprint,superscriptaddress,amsmath,amssymb,showpacs]{revtex4-1}
\usepackage{dcolumn}
\usepackage{graphicx}
\usepackage{float}
\usepackage{physics}
\usepackage[colorlinks=true,allcolors=blue]{hyperref}
\usepackage{mathrsfs}
\usepackage{inputenc}
\begin{document}

\title{Heavy quarkonium spectral function in the spinning black hole background}

\author{Zhou-Run Zhu }
\email{zhuzhourun@zknu.edu.cn}
\affiliation{School of Physics and Telecommunications Engineering, Zhoukou Normal University, Zhoukou 466001, China}

\author{Manman Sun}
\email{sunmm@zknu.edu.cn}
\affiliation{School of Physics and Telecommunications Engineering, Zhoukou Normal University, Zhoukou 466001, China}

\author{Rui Zhou}
\email{ruychou@zknu.edu.cn}
\affiliation{School of Physics and Telecommunications Engineering, Zhoukou Normal University, Zhoukou 466001, China}

\author{Zhuang Ma}
\email{mzhuang@zknu.edu.cn}
\affiliation{School of Physics and Telecommunications Engineering, Zhoukou Normal University, Zhoukou 466001, China}

\author{Jinzhong Han}
\email{hanjinzhong@zknu.edu.cn}
\affiliation{School of Physics and Telecommunications Engineering, Zhoukou Normal University, Zhoukou 466001, China}

\begin{abstract}
In this paper, we study the dissociation of heavy quarkonium in the spinning black hole background. Specifically, we analyze the spectral function of charmonium and bottomonium in the spinning black hole background and examine how the angular momentum affects the dissociation of $J/\Psi$ and $\Upsilon(1S)$. From the results, we find that the angular momentum and temperature decreases the peak height and expands the peak width of the spectral function, thereby enhancing the dissociation of heavy vector mesons. Moreover, the angular momentum has a stronger dissociation effect in the transverse orientation, revealing the directional influence of angular momentum.
\end{abstract}

\maketitle

\section{Introduction}\label{sec:01_intro}

It is believed that heavy-ion collisions under extreme conditions have created the strongly coupled quark-gluon plasma (QGP) and provide insights into the nature of QGP \cite{Arsene:2004fa,Adcox:2004mh,Back:2004je,Adams:2005dq}. Heavy quarkonium (such as $J/\Psi$ and $\Upsilon(1S)$) consists of a heavy quark and antiquark. One significant piece of evidence for the formation of QGP is the suppression of heavy quarkonium due to Debye screening \cite{Matsui:1986dk}. Heavy quarkonium influences the final yields and spectra since it strongly interacts with QGP. Heavy quarkonium mesons could be seen as a unique probe to investigate the properties of QGP. The melting or dissociation of heavy quarkonium can be related to the disappearance of particle peaks in the spectral function. The spectral function offers a deep understanding of the dissociation effects and behavior of heavy quarkonium states.

AdS/CFT correspondence \cite{Maldacena:1997re,Witten:1998qj,Gubser:1998bc} is a valuable tool for studying the properties of strongly coupled plasma. From the holographic perspective, the
gauge theory living on the boundary is dual to the gravitational theory in AdS space. This non-perturbative method has provided significant inspirations into the heavy flavor physics. The effect of temperature on scalar glueballs and scalar mesons has been studied in \cite{Colangelo:2009ra,Miranda:2009uw}. The studies of \cite{Dudal:2014jfa,Braga:2018zlu,Zhao:2021ogc} examine the effect of magnetic field on the dissociation of heavy quarkonium mesons. The meson dissociation phenomenon can be related to the suppression of the peak height of the spectral function. The influences of temperature and chemical potential on the melting of heavy quarkonium have been discussed in \cite{Mamani:2013ssa,Braga:2016wkm,Braga:2017oqw,Braga:2017bml,Mamani:2022qnf}. The researches of \cite{Braga:2015jca,Braga:2015lck} identify the masses and decay constants of the heavy quarkonium. The quasinormal modes of heavy quarkonium have been discussed in the holographic QCD models \cite{Braga:2019yeh,Braga:2019xwl,Zhao:2023yry}. In \cite{Zhao:2023pne,Braga:2023fac}, the dissociation of heavy quarkonium mesons in the rotating plasma has been investigated. The dissociation effects of heavy quarkonium in the anisotropic background have been studied in \cite{Chang:2024ksq}. The spectral functions in Gauss-Bonnet gravity can be found in \cite{Zhu:2024ddp}.

In this paper, we want to explore the heavy quarkonium spectral function in the spinning black hole background. The main motivation is that the nonzero angular momentum is detected in non-central heavy ion collisions and a large portion of angular momentum can be transferred to the QGP medium \cite{Liang:2004ph,Becattini:2007sr,Baznat:2013zx,STAR:2017ckg,Jiang:2016woz}. The angular momentum may produce significant observable effects on the physical quantities. Such as confinement-deconfinement phase transition \cite{Chen:2020ath,Braga:2022yfe,Zhao:2022uxc}, energy loss \cite{Hou:2021own,Chen:2022obe,Chen:2023yug} and entropy of heavy quarkonium \cite{Wu:2022ufk}. Other significative researches are detailed in \cite{McInnes:2014haa,McInnes:2016dwk,Cai:2023cjl,Zhang:2023psy,Zhao:2024ipr}.

In \cite{Zhao:2023pne,Braga:2023fac}, the authors extend the static background to a rotating background by a local Lorentz transformation \cite{Erices:2017izj,BravoGaete:2017dso,Awad:2002cz} and analyze the spectral functions in rotating medium. However, this local rotating background, created by the local Lorentz transformation, only represents a small neighbourhood around a rotating radius and a domain range less than 2$\pi$ wide. Studying the heavy quarkonium behaviors in spinning black holes \cite{Hawking:1998kw,Gibbons:2004ai,Gibbons:2004js,Garbiso:2020puw} may be a meaningful work. In \cite{Garbiso:2020puw}, the authors consider the Myers-Perry black holes which are vacuum solutions under Einstein gravity. The boundary in Myers-Perry black holes space is compact and the dual gauge theory is living on $S^3 \times \mathbb{R}$. In further research, the dual field theory is in non-compact spacetime when considering the planar black brane in the large black holes (high-temperature case). Thus, the spinning QGP medium lives on flat space $\mathbb{R}^{3,1}$ in this case. Inspired by this, we want to investigate the spectral function and discuss the dissociation of heavy quarkonium in spinning black holes.

The paper is organized as follows. In Sec.~\ref{sec:02}, we introduce the spinning Myers-Perry black hole. In Sec.~\ref{sec:03}, we study the spectral function in spinning Myers-Perry black hole. The conclusion and discussion are given in Sec.~\ref{sec:04}.

\section{Spinning Myers-Perry black hole background}\label{sec:02}

Before studying the spectral function of charmonium and bottomonium, we briefly review the spinning black hole background written by Hawking et al. The metric is given by \cite{Hawking:1998kw}
 \begin{equation}
\label{eqc1}
\begin{split}
ds^{2}=&-\frac{\Delta}{\rho^{2}}(dt_{H}-\frac{a\sin^{2}\theta_{H}}{\Xi_{a}}d\phi_{H}-\frac{b\cos^{2}\theta_{H}}{\Xi_{b}}d\psi_{H})^{2}+\frac{\Delta_{\theta_{H}}\sin^{2}\theta_{H}}{\rho^{2}}(adt_{H}-\frac{r_{H}^{2}+a^{2}}{\Xi_{a}}d\phi_{H})^{2}\\
 &+\frac{\Delta_{\theta_{H}}\cos^{2}\theta_{H}}{\rho^{2}}(bdt_{H}-\frac{r_{H}^{2}+b^{2}}{\Xi_{b}}d\psi_{H})^{2}+\frac{\rho^{2}}{\Delta}dr_{H}^{2}-\frac{\rho^{2}}{\Delta_{\theta_{H}}}d\theta_{H}^{2}\\
 &+\frac{1+\frac{r_{H}^{2}}{L^{2}}}{r_{H}^{2}\rho^{2}}(abdt_{H}-\frac{b\left(r^{2}+a^{2}\right)\sin^{2}\theta_{H}}{\Xi_{a}}d\phi_{H}-\frac{a\left(r^{2}+b^{2}\right)\cos^{2}\theta_{H}}{\Xi_{b}}d\psi_{H})^{2},
 \end{split}
\end{equation}
with
 \begin{equation}
\label{eqc11}
\begin{split}
\Delta=\frac{1}{r_{H}^{2}}(r_{H}^{2}+a^{2})(r_{H}^{2}+b^{2})(1+\frac{r_{H}^{2}}{L^{2}})-2M,\\
 \Delta_{\theta_{H}}=1-\frac{a^{2}}{L^{2}}\cos^{2}\theta_{H}-\frac{b^{2}}{L^{2}}\sin^{2}\theta_{H},\\
 \rho=r_{H}^{2}+a^{2}\cos^{2}\theta_{H}+b^{2}\sin^{2}\theta_{H},\\
 \Xi_{a}=1-\frac{a^{2}}{L^{2}},\\
\Xi_{b}=1-\frac{b^{2}}{L^{2}},\\
 \end{split}
\end{equation}
where $t_H$, $L$ and $r_H$ denotes time, AdS radius and AdS radial coordinate respectively. $(\phi_H, \psi_H, \theta_H)$ represent angular Hopf coordinates. The angular momentum parameters are denoted by $a$ and $b$. In this work, we take consider spinning Myers-Perry black hole case, namely $a=b$ \cite{Gibbons:2004ai,Gibbons:2004js}.

To facilitate the analysis, more convenient coordinates can be adopted and the mass can be reparameterized \cite{Murata:2008xr}
 \begin{equation}
\label{eqc111}
\begin{split}
t=t_{H},\\
 r^{2}=\frac{a^{2}+r_{H}^{2}}{1-\frac{a^{2}}{L^{2}}},\\
\theta=2\theta_{H},\\
 \phi=\phi_{H}-\psi_{H},\\
\psi=-\frac{2at_{H}}{L^{2}}+\phi_{H}+\psi_{H},\\
b=a,\\
\mu=\frac{M}{(L^{2}-a^{2})^{3}}.\\
 \end{split}
\end{equation}

The metric (\ref{eqc1}) can be rewrite as
 \begin{equation}
\label{eqc2}
\begin{split}
ds^{2}=&-(1+\frac{r^{2}}{L^{2}})dt^{2}+\frac{dt^{2}}{G(\text{r)}}+\frac{r^{2}}{4}((\sigma^{1})^{2}+(\sigma^{2})^{2}+(\sigma^{3})^{2})+\frac{2\mu}{r^{2}}(dt+\frac{a}{2}\sigma^{3})^{2},
 \end{split}
\end{equation}
with
 \begin{equation}
\label{eqc3}
\begin{split}
G(r)=1+\frac{r^{2}}{L^{2}}-\frac{2\mu(1-\frac{a^{2}}{L^{2}})}{r^{2}}+\frac{2\mu a^{2}}{r^{4}},\\
\mu=\frac{r_{h}^{4}(L^{2}+r_{h}^{2})}{2L^{2}r_{h}^{2}-2a^{2}(L^{2}+r_{h}^{2})},\\
\sigma^{1}=-sin\psi dtd\theta+cos\psi sin\theta d\phi,\\
\sigma^{2}=cos\psi d\theta+sin\psi sin\theta d\phi,\\
\sigma^{3}=d\psi+cos\theta d\phi,\\
 \end{split}
\end{equation}
where
\begin{equation}
\label{eqc4}
\begin{split}
&-\infty<t<\infty,\ r_{h}<r<\infty,\ 0\leq\theta\leq\pi,\\
&0\leq\phi\leq2\pi,\ 0\leq\psi\leq4\pi.
 \end{split}
\end{equation}

In order to get the planar black brane, one can employ the coordinate transformation as \cite{Garbiso:2020puw}
 \begin{equation}
\label{eqc5}
\begin{split}
t=\tau,\\
\frac{L}{2}(\phi-\pi)=x_{1},\\
\frac{L}{2}tan(\theta-\frac{\pi}{2})=x_{2},\\
\frac{L}{2}(\psi-2\pi)=x_{3},\\
r=\tilde{r},\\
 \end{split}
\end{equation}
where $(\tau, \widetilde{r}, x, y, z)$ are new coordinates. In the next step, the coordinates can be scaled with a power of scaling factor $\beta$
\begin{equation}
\label{eqc6}
\begin{split}
\tau\rightarrow\beta^{-1}\tau,\\
x_{1}\rightarrow\beta^{-1}x_{1},\\
x_{2}\rightarrow\beta^{-1}x_{2},\\
x_{3}\rightarrow\beta^{-1}x_{3},\\
\tilde{r}\rightarrow\beta\tilde{r},\\
\tilde{r_{h}}\rightarrow\beta\tilde{r_{h}}.(\beta\rightarrow \infty)\\
 \end{split}
\end{equation}

Then one can obtain the Schwarzschild black brane metric which is boosted in $\tau- x_3$ plane \cite{Garbiso:2020puw}
\begin{equation}
\label{eqc7}
\begin{split}
ds^{2}=&\frac{r^{2}}{L^{2}}(-d\tau^{2}+dx_{1}^{2}+dx_{2}^{2}+dx_{3}^{2}+\frac{r_{h}^{4}}{r^{4}(1-\frac{a^{2}}{L^{2}})}(d\tau+\frac{a}{L}dx_{3})^{2})+\frac{L^{2}r^{2}}{r^{4}-r_{h}^{4}}dr^{2},
 \end{split}
\end{equation}
where Schwarzschild black brane is recovered when $a=0$.

In order to calculate conveniently, we replace the holographic fifth coordinate $r$ by $1/z$ ($r=1/z$)
\begin{equation}
\label{eqc8}
\begin{split}
ds^{2}=& \frac{1}{z^{2}L^{2}}[-d\tau^{2}+dx_{1}^{2}+dx_{2}^{2}+dx_{3}^{2}+\frac{z^{4}}{z_{h}^{4}(1-\frac{a^{2}}{L^{2}})}(d\tau+\frac{a}{L}dx_{3})^{2}]+\frac{L^{2}z_{h}^{4}}{z^{2}(z_{h}^{4}-z^{4})}dz^{2}.
 \end{split}
\end{equation}

The temperature can be given by\cite{Garbiso:2020puw}
\begin{equation}
\label{eqb1}
T=\frac{\sqrt{L^{2}-a^{2}}}{z_{h}\pi L^{3}},
\end{equation}
where $z_h$ denotes the horizon. From the perspective of the temperature, the values of a should satisfy the condition $a<L$ to ensure the temperature is positive.

\section{Spectral function in the spinning Myers-Perry black hole background}\label{sec:03}

In this section, we will study the spectral function of charmonium and bottomonium by a phenomenological model proposed in Ref.\cite{Braga:2018zlu}. One can derive the formulas of spectral function by following Ref.\cite{Braga:2018zlu}. We investigate the spectral function in the spinning Myers-Perry black hole background. In order to facilitate calculation, we rewrite the metric (\ref{eqc8}) with a general form,
\begin{equation}
\begin{split}
\label{eqa1}
ds^{2}=& -g_{\tau \tau} d\tau^2 + g_{xx_{1}} (dx_{1}^{2}+ dx_{2}^{2})+g_{xx_{3}}dx_{3}^{2}+g_{\tau x_3}d\tau dx_3 + g_{x_3 \tau}dx_3d\tau+ g_{zz} dz^{2}.
 \end{split}
\end{equation}

The vector field $V_m =(V_\mu, V_z)$ denotes the heavy quarkonium and corresponds to gauge theory current $J^\mu=\overline{\Psi} \gamma^\mu \Psi$. The bulk action is \cite{Braga:2018zlu}

\begin{equation}\label{eq:x1}
I= -\int d^{4}x dz \frac{Q}{4}F_{mn}F^{mn},
\end{equation}
where $F_{mn}=\partial_m V_n -\partial_n V_m$ and $Q=\frac{\sqrt{-g}}{e^{\phi(z)}g^2_5}$. The dilaton field $\phi(z)$ is used to characterise vector mesons\cite{Braga:2018zlu},
\begin{equation}\label{eq:x4}
\phi(z)= w^2 z^2+Mz+\tanh(\frac{1}{Mz}-\frac{w}{\sqrt{\Gamma}}),
\end{equation}
where $w$ represents quark mass, $\Gamma$ is string tension and $M$ denotes non-hadronic decay. We consider that the metric (\ref{eqc8}) is not affected by the dilaton background. The energy parameters ($w$, $\Gamma$ and $M$) are determined by fitting the masses spectrum. The specific values of the energy parameters for charmonium and bottomonium are \cite{Braga:2018zlu}
\begin{equation}
\label{eqav5}
\begin{split}
&  w_c=1.2GeV, \sqrt{\Gamma_c}=0.55GeV, M_c = 2.2GeV;\\
 & w_b=2.45GeV, \sqrt{\Gamma_b}=1.55GeV, M_b = 6.2GeV.
 \end{split}
\end{equation}

The spectral function can be calculated by membrane paradigm \cite{Iqbal:2008by}. The equation of motion can be obtained from Eq.(\ref{eq:x1})
\begin{equation}
\label{eqa8}
 \partial^m (Q F_{mn})=0 ,
\end{equation}
and the conjugate momentum of the gauge field for z-foliation
\begin{equation}
\label{eqa9}
 j^\mu= -Q F^{z\mu}.
\end{equation}

One can assume that the plane wave solution for vector field is propagating in $x_3$ direction. Then we can consider the equations of motion (\ref{eqa8}) have longitudinal and transverse channels. The longitudinal fluctuations are along ($\tau, x_3$) and transverse fluctuations are along ($x_1, x_2$). Therefore, the longitudinal components of Eq.(\ref{eqa8}) are
\begin{equation}
\label{eqav8}
\begin{split}
&-\partial_z j^\tau-Q(g^{xx_3}g^{\tau \tau}+g^{x_3 \tau}g^{\tau x_3 })\partial_{x_3}F_{x_3 \tau}=0,\\
& -\partial_z j^{x_3}+ Q(g^{\tau \tau}g^{xx_3}+g^{\tau x_3 }g^{x_3 \tau})\partial_{\tau}F_{x_3 \tau}=0,\\
& \partial_{x_3} j^{x_3}+\partial_\tau j^\tau =0.
\end{split}
\end{equation}

Adopting the Bianchi identity, one can obtain
\begin{equation}
\label{eqa91}
\partial_z F_{x_3 \tau}-\frac{g_{zz}}{Q}\partial_\tau[g_{xx_3}j^{x_3}+g_{x_3 \tau}j^{\tau}]-\frac{g_{zz}}{Q}\partial_{x_3}[g_{\tau \tau}j^\tau-g_{\tau x_3 }j^{x_3}]=0.
\end{equation}

One can get the conductivity of longitudinal channel as
\begin{equation}
\label{eqav9}
\begin{split}
& \sigma_L (\omega,\overrightarrow{p},z)=\frac{j^{x_3}(\omega,\overrightarrow{p},z)}{F_{x_3 \tau}(\omega,\overrightarrow{p},z)},\\
& \partial_z \sigma_L=\frac{\partial_z j^{x_3}}{F_{x_3 \tau}}-\frac{ j^{x_3}}{F^2_{x_3 \tau}}\partial_z F_{x_3 \tau}.
 \end{split}
\end{equation}

Using Kubo formula, one can get the relations between the longitudinal AC conductivity $\sigma_L$ and the retarded Green's function
\begin{equation}
\label{eqa12}
 \sigma_L(\omega)=-\frac{G^L_R (\omega)}{i \omega}.
\end{equation}

Combining Eq.(\ref{eqav8}), (\ref{eqa91}) with momentum limit $P$ = ($\omega$, 0, 0, 0), one can the flow equation of longitudinal channel
\begin{equation}
\label{eqav92}
 \partial_z \sigma_L=\frac{i\omega g_{xx_3}g_{zz}}{Q}(\sigma^2_L-\frac{Q^2(g^{\tau \tau}g^{xx_3}+g^{\tau x_3}g^{x_3 \tau})}{g_{xx_3}g_{zz}}).
\end{equation}

The flow equation of transverse channel can be obtained from similar process
\begin{equation}
\label{eqav92}
 \partial_z \sigma_T=\frac{i\omega g_{xx_1}g_{zz}}{Q}(\sigma^2_T-\frac{Q^2 g^{zz}g^{\tau \tau}}{g^2_{xx_1}}).
\end{equation}

Utilizing the regularity condition at horizon, the equation can be solved with $\partial_z \sigma_{L/T}(\omega,z)=0$. Then one can get the expression of spectral function
\begin{equation}
\label{eqa131}
 \rho(\omega)\equiv -Im G_R (\omega)=\omega Re \sigma(\omega,0).
\end{equation}

\begin{figure}[H]
    \centering
      \setlength{\abovecaptionskip}{0.1cm}
    \includegraphics[width=11cm]{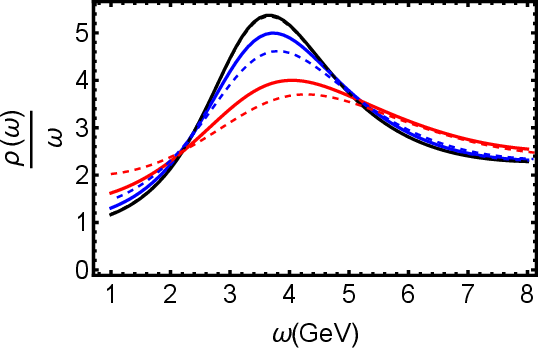}
    \caption{\label{fig1} Spectral functions for charmonium at $T=0.6$ GeV for different values of angular momentum $a$. The black, blue and red line denote $a = 0.1,\ 0.4$, and $0.7$, respectively. The solid line (dashed line) represents the longitudinal (transverse) channel.}
\end{figure}

In the limitation of the large black hole, the angular velocity $\Omega = a/L^2$ \cite{Garbiso:2020puw}. The linear velocity is the product of the angular velocity and the radius of rotation. However, the spinning Myers-Perry black hole (angular momentum parameters $a=b$) is peculiar. From the results of the integral curves of the axial angular momentum Killing vector field, it can be found that there is no axis of rotation in the Myers-Perry black hole \cite{Amano:2023bhg}. One can consider that the fluid flow lines for an embedding in $\mathbb{R}^{3}$ lies on a torus \cite{Amano:2023bhg}. Moreover, The ultra-local behavior of fluid flow can be considered as a uniform motion, which involved is that of uniformly rotating circles of a Hopf fibration of the 3-sphere constituting the spatial part of the four-dimensional spacetime \cite{Amano:2023bhg}. The induced rotation has no axis in the dual field theory, thus it is hard to calculate the linear velocity. However, we can discuss the values of angular momentum $a$ from other certain restrictions. The background will rotate at the speed of light when one or both of $\Xi_{a}$ and $\Xi_{b}$ vanish ($a=L$ or/and $b=L$) \cite{Hawking:1998kw}. In order to ensure the temperature is positive, the values of a should satisfy the condition $a<L$. From the results of the momentum diffusion coefficient, one can find that the spinning black brane is stable when $a<0.75L$ \cite{Garbiso:2020puw}. Therefore, it is advisable to set $a<0.75L$ in the numerical calculations.

In Ref. \cite{Amano:2023bhg}, the authors investigate the hydrodynamics of the Myers-Perry black hole. They discover that low energy fluctuations treat rotating fluid as a boost with the boost parameter $a$. The frequency and momentum of the collective excitations around rotating states are small compared to the temperature and angular momentum scale. This allows us to treat the rotating fluid as a boosted fluid on its relative scale. This finding suggests that at intermediate to high temperatures, the plasma may be well approximated by models of boosted fluids. The rotation has no axis in the dual field theory. Additionally, the behavior of fluid flow is uniform motion, as indicated by the results of the hydrodynamic modes, Lyapunov exponents, and butterfly velocities. Thus, the fluid flow for an embedding in $\mathbb{R}^{3}$ is lying on a torus and is uniform motion in the Myers-Perry black hole. Therefore, we consider that the dissociation effect homogeneous in this peculiar and spinning Myers-Perry black hole.

In order to study how the angular momentum affects the dissociation of heavy vector mesons, we calculate the spectral functions in the spinning black hole background. In the calculations, we set $L=1$ for simplicity. We investigate the dissociation behaviors with high temperature since the background geometry performs a planar limit black brane in large black hole (high-temperature case) \cite{Garbiso:2020puw}. Additionally, the spinning black brane is stable when $a < 0.75L$ and we will take $a=0.1, 0.4, 0.7$ in the calculations.

In Fig.~\ref{fig1}, we display the spectral functions for charmonium at $T=0.6$ GeV for different values of angular momentum $a$. The bell shape (peak feature) describes the quasiparticle state. The spectral function of charmonium may contain a set of peaks. The first, the second, the third peak corresponds to the 1S state ($J/\Psi$), 2S state, 3S state, respectively and so one. We focus on the influence of angular momentum and temperature on the 1S state ($J/\Psi$). The width of spectra is inversely proportional to decay rate which is related to the stability. The results show that increasing the angular momentum suppresses the height and broadens the peak width in parallel and transverse channels. This phenomenon implies that the angular momentum enhances the dissociation effect of $J/\Psi$. Moreover, it is evident that the angular momentum has a stronger dissociation effect in the transverse case, revealing the directional influence of angular momentum.

\begin{figure}[H]
    \centering
      \setlength{\abovecaptionskip}{0.1cm}
    \includegraphics[width=11cm]{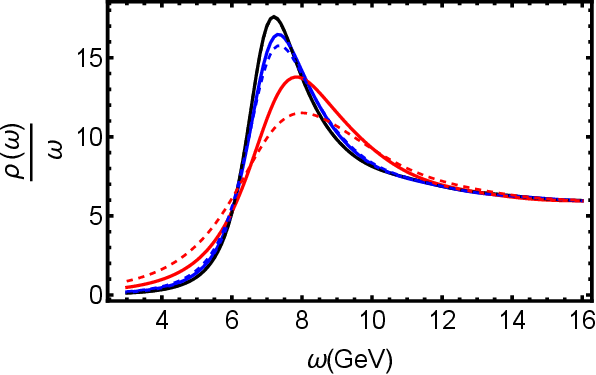}
    \caption{\label{fig2} Spectral functions for bottomonium at $T=0.6$ GeV for different values of angular momentum $a$. The black, blue and red line denote $a = 0.1,\ 0.4$, and $0.7$, respectively. The solid line (dashed line) represents the longitudinal (transverse) channel.}
\end{figure}

In Fig.~\ref{fig2}, we plot the spectral functions for bottomonium at $T=0.6$ GeV for different values of angular momentum $a$. The spectral function of bottomonium also contain a set of peaks and we focus on the influence of angular momentum and temperature on the 1S state ($\Upsilon(1S)$). It is obviously that the angular momentum decreases the height and enlarges the peak width of spectra, indicating the angular momentum favors the dissociation of $\Upsilon(1S)$. Furthermore, we observe that the angular momentum has a stronger dissociation effect in the transverse case. From the results of Figs.~\ref{fig1} and ~\ref{fig2}, one can summarize that the angular momentum promotes the dissociation of heavy quarkonium. This may be due to the angular momentum decreases the screening length of quark-antiquark pair, hence increases the quarkonium dissociation \cite{Zhu:2024dwx}. Another physical interpretation is the angular momentum enhances the entropic force of quark-antiquark pair, thereby enhancing the dissociation \cite{Zhu:2024dwx}.

\begin{figure}[H]
    \centering
      \setlength{\abovecaptionskip}{0.1cm}
    \includegraphics[width=11cm]{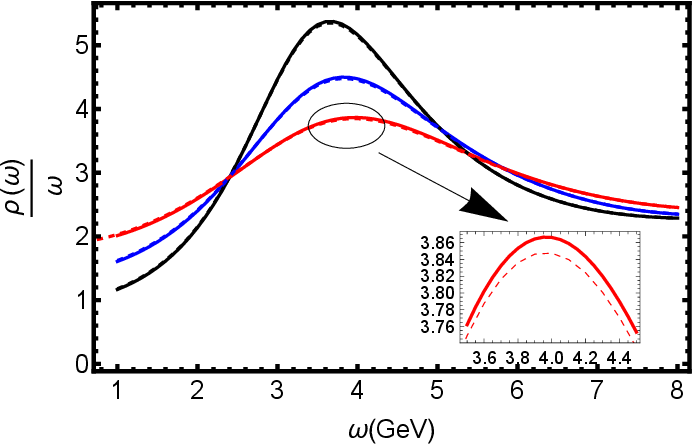}
    \caption{\label{fig3} Spectral functions for charmonium at $a=0.1$ for different values of temperature $T$. The black, blue and red line denote $T = 0.6,\ 0.7$, and $0.8$ GeV, respectively. The solid line (dashed line) represents the longitudinal (transverse) channel.}
\end{figure}

\begin{figure}[H]
    \centering
      \setlength{\abovecaptionskip}{0.1cm}
    \includegraphics[width=11cm]{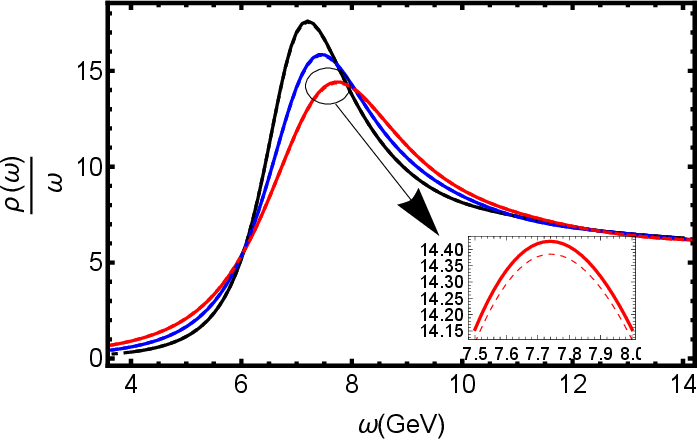}
    \caption{\label{fig4} Spectral functions for bottomonium at $a=0.1$ for different values of temperature $T$. The black, blue and red line denote $T = 0.6,\ 0.7$, and $0.8$ GeV, respectively. The solid line (dashed line) represents the longitudinal (transverse) channel.}
\end{figure}

In Figs.~\ref{fig3} and ~\ref{fig4}, we draw the spectral functions for charmonium and bottomonium at $a=0.1$ for different values of temperature, respectively. One can observe that the temperature decreases the height and enlarges the peak width in parallel and transverse cases, revealing the temperature accelerates the dissociation of the heavy quarkonium. Furthermore, the angular momentum has a stronger dissociation effect in the transverse orientation.

\section{Conclusion and discussion}\label{sec:04}

In this paper, we want to explore the dissociation of heavy quarkonium in the spinning black hole background. The nonzero angular momentum is detected in non-central heavy ion collisions and may produce significant observable effects on the heavy quarkonium. To be specific, we calculate the spectral functions and discuss the effect of angular momentum on the dissociation of $J/\Psi$ and $\Upsilon(1S)$.

The results reveal that the angular momentum decreases the height and enlarges the peak width of spectral function, which indicates the angular momentum favors dissociation of $J/\Psi$ and $\Upsilon(1S)$. Moreover, the angular momentum has a stronger dissociation effect in transverse case. Additionally, the temperature decreases the height and enlarges the peak width in parallel and transverse cases, revealing the temperature accelerates the dissociation of the heavy quarkonium.

We expect that the results of the spectral functions in the spinning black hole background could provide more insight into the investigation of the dissociation of heavy quarkonium. Finally, it is worth exploring the configuration entropy in spinning black hole background. We leave this work in the future.

\section*{Acknowledgments}

The authors would like to acknowledge Yan-Qing Zhao for helpful discussion. Jinzhong Han is supported by the Foundation of the Natural Science Foundation of Henan Province under Grant No.242300421398. Rui Zhou is supported by the Science and Technology Development Plan Project of Henan Province No. 242102230085. Rui Zhou is also supported by the National Natural Science Foundation of China under Grant No. 12404470. Manman Sun is supported by the National Natural Science Foundation of China under Grant No.12305076. Zhou-Run Zhu is supported by the High Level Talents Research and Startup Foundation Projects for Doctors of Zhoukou Normal University No. ZKNUC2023018.

\section*{Data Availability Statement}

This manuscript has no associated data or the data will not be deposited.

\end{document}